\documentclass[twocolumn,secnumarabic,amssymb, nofootinbib, nobibnotes,superscriptaddress,showkeys, aps, prd]{revtex4-2}
\usepackage{graphicx}	
\usepackage{amsmath}	
\usepackage{amssymb}	
\usepackage{hyperref}
\usepackage{subfigure}
\usepackage{romannum}

\usepackage{color}

\begin{document}

\title{New Constraints on Lorentz Invariance Violation at High Redshifts from Multiband of GRBs}

\author{Mingyue Chen}
\thanks{These authors contributed equally to this work.}
\affiliation{School of Electronic Science and Engineering, Chongqing University of Posts and Telecommunications, Chongqing 400065, China}

\author{Jun Tian}
\thanks{These authors contributed equally to this work.}
\affiliation{School of Electronic Science and Engineering, Chongqing University of Posts and Telecommunications, Chongqing 400065, China}
\affiliation{School of Physics and Astronomy, China West Normal University, Nanchong 637002, China}

\author{Yu Pan}
\email{panyu@cqupt.edu.cn}
\affiliation{School of Electronic Science and Engineering, Chongqing University of Posts and Telecommunications, Chongqing 400065, China}
 
\author{Tonghua Liu}
\affiliation{School of Physics and Optoelectronic Engineering, Yangtze University, Jingzhou, 434023, China.}

\author{Shuo Cao}
\email{caoshuo@bnu.edu.cn}
\affiliation{Department of Astronomy, Beijing Normal University, Beijing 100875, China.}

\begin{abstract}
In the gravity quantum theory, the quantization of spacetime may lead to the modification of the dispersion relation between the energy and the momentum and the Lorentz invariance violation (LIV). High energy and long-distance gamma-ray bursts (GRBs) observations in the universe provide a unique opportunity to test the possibility of LIV. In this work, we use 88 time delays from GRBs ($0.117 < z < 6.29$), and provide a cosmological model-independent approach based on the luminosity distance data from 174 GRBs to test LIV. Combining the observation data from multiband of GRBs provides us with an opportunity to mitigate the potential systematic errors arising from variations in the physical characteristics among diverse object populations, and to add a higher redshift dataset for testing the energy-dependent velocity caused by the corrected dispersion relationship of photons. These robust limits of the energy scale for the linear and quadratic LIV effects are $E_{\mathrm{QG},1} \ge 1.5\times 10^{15}$ GeV, and $E_{\mathrm{QG},2} \ge 8.5\times 10^{9}$ GeV, respectively. It exhibits a significantly reduced value compared to the energy scale of Planck in both scenarios of linear and quadratic LIV. 
\end{abstract}

\maketitle

\section{\label{sec:level1}Introduction}

To better explain the microstructure of spacetime, quantum gravity theory combines relativity with quantum mechanics~\cite{Amelino-Camelia:2008aez}.
The theory of quantum gravity states that spacetime exhibits a discontinuous foam structure at energies near the Planck scale, $E_{\mathrm{QG}}\approx E_{\mathrm{Pl}} = \sqrt{\hbar c^{5}/G}\simeq 1.22\times 10^{19}$ GeV, and that the foam structure could potentially be subject to interaction from high-energy photons~\cite{Amelino-Camelia:1996bln, Mattingly:2005re}.
Lorentz invariance, a fundamental theorem in special relativity, states the preservation of physical laws under Lorentz transformations, unveiling the inherent symmetry of spacetime.
However, the interaction of the spacetime foam structure with high-energy photons causes the speed of light to change, which is contrary to Lorentz invariance, a phenomenon known as Lorentz invariance violation (LIV)~\cite{2004PhLB..585....1M}.
Hence, the velocities of photons with different energies exhibit a bias, indicating that photons emitted from the same source but with varying energies do not arrive at the observer simultaneously~\cite{Ellis:2002in,Ellis:2005sjy}.
Measurement of the speed of light in a vacuum will help us to test the LIV.
Promising probes include gravitational waves, ultra-high-energy cosmic rays, and neutrinos~\cite{Addazi:2021xuf, Vasileiou:2013vra, 2015APh....61..108Z}.
Among these, gamma-ray bursts (GRBs) are particularly compelling: as the most luminous and distant electromagnetic transients in the universe, they emit photons over a broad energy range on short timescales, making them ideal laboratories for testing LIV effects~\cite{ Pan:2015cqa, Wei:2016exb,Liao:2022gde}.

GRBs are usually classified into two types based on their durations: long and short GRBs~\cite{Kouveliotou:1993yx}. Long GRBs are typically related to core-collapse supernovae (SNe)~\cite{Woosley:2006fn}, while short GRBs are related to neutron star mergers~\cite{Nakar:2007yr}. In addition to the two types of GRBs mentioned above, several studies have proposed other subcategories ~\cite{Kulkarni:2016onj,Bhave:2017nnx}.
The search for LIV signatures in GRB data has evolved significantly with advances in GRB instrumentation. 
Early efforts to constrain LIV relied on GRB observations from instruments such as the Burst and Transient Source Experiment (BATSE), the Reuven Ramaty High Energy Solar Spectroscopic Imager (RHESSI), the Swift Gamma-Ray Burst Mission (Swift), the Konus instrument aboard the Wind spacecraft (Konus-Wind), and the High Energy Transient Explorer 2 (HETE-2)~\cite{RodriquezMartinez:2006xc,Bolmont:2006kk,Lamon:2007wr}. A major breakthrough came with the launch of the Fermi satellite, whose Large Area Telescope (LAT) provided exceptional sensitivity to high-energy GRB emission, reaching up to tens of GeV. Using this capability, the Fermi Collaboration obtained significantly tighter LIV constraints from the analyzes of GRB 080916C~\cite{Fermi-LAT:2009owx} and GRB 090510~\cite{FermiGBMLAT:2009nfe}, with further independent results reported for GRB 090510~\cite{Xiao:2009xe}.

However, most studies constrain LIV within the framework of standard cosmological models, such as the $\Lambda$CDM model~\cite{2024rpgt.book..433D}. These models are derived under the assumption of exact Lorentz invariance and do not incorporate potential LIV effects on cosmic expansion or photon propagation. This reliance on Lorentz-invariant cosmology introduces a conceptual inconsistency, motivating the development of model-independent approaches to LIV constraints. Several such model-independent methods have been proposed. One approach parameterizes the luminosity distance using a Taylor series expansion, offering a general description of cosmic expansion and thereby enabling a more robust estimation of LIV-induced time delays without assuming a specific cosmological model~\cite{Zou:2017ksd}. However, it has been noted that this Taylor expansion becomes rapidly divergent at higher redshifts, limiting its applicability~\cite{2007gr.qc.....3122C,2005GReGr..37.1555T}.
An alternative method employs Gaussian Processes (GP) to reconstruct the cosmic expansion history $H(z)$ directly from observational data~\cite{Pan:2020zbl}. 
This reconstructed $ H(z) $ is subsequently combined with GRB time delay measurements in the redshift range $ 0.165 < z < 2.5 $ to constrain the energy-dependent dispersion relation of high-energy photons.  
Although the full GRB time delay sample extends to redshifts as high, events with $ z > 2.5 $ are excluded from the analysis because the GP-reconstructed $ H(z) $ is not reliably constrained beyond the maximum redshift of the available $ H(z) $ measurements.
Due to uncertainties in the emission mechanisms of GRBs, larger samples are required to improve the statistical robustness of LIV constraints. Therefore, a model-independent method capable of operating over a wide redshift range is needed to match the extensive redshift coverage of GRBs.

In this study, we present a novel model-independent approach to constrain LIV over an extended redshift range, reaching into the high-redshift range of GRBs.
Our method is based entirely on GRB data. First, we reconstruct the cosmic expansion history using luminosity distances from 174 GRBs, spanning a redshift range of $0.117 < z < 9.4$~\cite{Tang:2021elg}. Second, we used a sample of 88 time delays from long GRBs, covering $0.117 \leq z \leq 6.29$~\cite{Wei:2016exb,2006APh....25..402E,Xiao:2022ovb}. Crucially, both the expansion history and the time-delay data employed in this work are derived exclusively from GRBs. That could potentially mitigate systematic errors arising from variations in the physical characteristics among diverse object populations. Moreover, the wide redshift coverage of the GRB luminosity distance sample enables the inclusion of a significantly larger set of high-redshift time-delay events. In particular, our analysis extends to $z = 6.29$, surpassing the redshift reach of previous model-independent studies, such as the GP approach of Pan et al.~\cite{Pan:2020zbl}.

This paper is structured as follows. 
In Section~\ref{sec:leve2}, we present a simple derivation of the time delay caused by LIV. 
Section~\ref{sec:leve3} describes the methods and datasets used in this study, including time delay data, luminosity distance data, and the Gaussian Process regression reconstruction method. 
In Section~\ref{sec:results}, we present our main results. 
Finally, Section~\ref{sec:conclusion} provides a summary and concluding remarks.

\section{\label{sec:leve2}Time Delay of Lorentz Invariance Violation}

An important observational feature of GRBs is the spectral lag, which refers to a time delay in the arrival of photons with different energies at Earth. The time delay between high- and low-energy photons is defined as~\cite{1997ApJ...486..928B}
\begin{eqnarray}
\Delta t = t_{\mathrm{low}} - t_{\mathrm{high}},
\label{eq:1}
\end{eqnarray}
where $t_{\mathrm{high}}$ and $t_{\mathrm{low}}$ are the arrival times on Earth of photons with high and low energies, respectively. A positive $\Delta t$ implies that high-energy photons arrive earlier, while a negative value indicates that low-energy photons arrive earlier. This energy-dependent arrival time could, in principle, arise from intrinsic emission mechanisms within the source. However, it may also carry signatures of new physics during photon propagation over cosmological distances, particularly from effects predicted by quantum gravity theories.

Quantum gravity models suggest that Lorentz invariance may be broken at the Planck energy scale $E_{\mathrm{QG}} \sim 10^{19}$ GeV. One possible manifestation of this violation is an energy-dependent modification of the photon propagation speed. The modified dispersion relation can be written as
\begin{eqnarray}
E^{2} = p^{2} c^{2} \left[1 - s_{\pm} \left(\frac{E}{E_{\mathrm{QG}}}\right)^{n}\right],
\label{eq:2}
\end{eqnarray}
where $E$ and $p$ denote the energy and momentum of the photon, respectively.
The variables $s_{\pm}$ take values of $\pm 1$, and when $s_{\pm} = 1$, the speed of light decreases with increasing energy, corresponding to subluminal LIV; when $s_{\pm} = -1$, the speed of light increases with energy, indicating superluminal LIV.
In this work, we adopt $s_{\pm} = 1$, as most quantum gravity models predict that higher-energy photons propagate more slowly than lower-energy ones~\cite{1995AnPhy.243...90L,1997IJMPA..12..607A}.
The parameter $n$ represents the order of the vacuum dispersion relation, when $n$ is equal to 1 or 2, it represents linear and quadratic models, respectively.

Assuming that the photon group velocity is still given by $v=dE/dp$, the photon velocity becomes
\begin{eqnarray}
v(E) = c \left[1 - s_{\pm} \frac{n+1}{2} \left(\frac{E}{E_{\mathrm{QG}}}\right)^{n}\right].
\label{eq:3}
\end{eqnarray}

Over cosmological distances, this tiny velocity difference accumulates into a measurable time delay. The LIV-induced delay between photons of observed energies $E$ and $E_0$ is given by~\cite{Jacob:2008bw,Ellis:2005sjy}
\begin{eqnarray}
\Delta t_{\mathrm{LIV}} = - \frac{1+n}{2H_{0}}
\frac{E^{n} - E_{0}^{n}}{E_{\mathrm{QG},n}^{n}}
\int_{0}^{z} \frac{(1+z')^{n}}{h(z')}  dz',
\label{eq:4}
\end{eqnarray}
where $H(z)$ is the Hubble parameter, $H_{0}$ is the Hubble constant, and $h(z)=H(z)/H_{0}$.
We define the dimensionless function
\begin{eqnarray}
K(z) = \int_{0}^{z} \frac{(1+z')^{n}}{h(z')} dz',
\label{eq:5}
\end{eqnarray}
the $K(z)$ is the term in the time delay expression $(\Delta t_{\mathrm{LIV}})$ induced by LIV associated with the cosmological model. In this paper, we reconstruct $K(z)$ from the luminosity distance of the calibrated 174 GRBs, avoiding the use of cosmological models.

In addition to the propagation effect, photons of different energies may be emitted at different times within the GRB source itself, producing an intrinsic time delay. Some studies have pointed out that multi-energy photons are emitted at different times from the source~\cite{2000ApJ...534..248N,2010arXiv1003.0229U,2016A&A...592A..95M}. However, the intrinsic delay cannot yet be described by a universally accepted quantitative model due to the complexity of GRB emission mechanisms.
In the early analyses, the intrinsic time delay was typically assumed to be a constant for all GRBs~\cite{Ellis:2005sjy}. Later studies proposed a more physically motivated power-law form, in which the intrinsic lag depends explicitly on the photon energy~\cite{Wei:2016exb}. The power-law model is
\begin{eqnarray}
	\Delta t_{\mathrm{int}} = \tau \left[ (\frac{E}{\mathrm{keV}})^{\alpha} - (\frac{E_{0}}{\mathrm{keV}})^{\alpha} \right]
	\label{eq:6}.
\end{eqnarray}
The median value of the lowest energy band, denoted as $E_{0}$, is considered along with two free parameters $\tau$ and $\alpha$.

The total observed time delay between photons of different energies can therefore be written as the sum of the LIV-induced and intrinsic contributions
\begin{eqnarray}
\Delta t = \Delta t_{\mathrm{LIV}} + \Delta t_{\mathrm{int}}(1+z),
\label{eq:7}
\end{eqnarray}
where the factor $(1+z)$ is the cosmological time-dilation effect.

In this framework, the absence of a statistically significant LIV signal allows us to place a robust lower limit on the quantum gravity energy scale $E_{\mathrm{QG},n}$.

\section{\label{sec:leve3}Data-set of GRBs and Model-independent methods}
\subsection{\label{sec:leve31}Data-set of GRBs}

In this paper, we implement cosmological model-independent constraints for LIV via multiband observations of long GRBs. First, we collect a sample of 88 time delays from long GRBs. Later, we reconstruct $K(z)$ from the luminosity distance data of 174 GRBs calibrated by Tang et al.~\cite{Tang:2021elg}, using a cosmological model-independent method.

In this work, we utilize 88 time delays from long GRBs, spanning a readshift range of $0.017<z<6.29$ and energy bands from $10$ keV to $20$ MeV. The dataset comprises three distinct subsets.
The first subset includes 37 time delays from the long GRB 160625B, reported in Table~1 of Wei et al.~\cite{Wei:2016exb}.  This burst was observed by the Fermi Gamma-ray Burst Monitor (GBM) and the LAT. It comprises three distinct emission episodes and has a total duration of approximately 770 seconds~\cite{Zhang:2016poo}. Its brightness during the second burst was of such intensity that it facilitated the extraction of its light curves across various energy ranges~\cite{2016GCN.19581....1B, 2016GCN.19580....1D}. These time delays were derived using the cross-correlation function (CCF) method~\cite{2012ApJ...748..132Z}, between the $10 \sim 12$ keV light curve and those in higher-energy bands ranging from 15 keV to 20 MeV. The second subset consists of 30 time delays from long GRBs selected from the mixed sample in Table~1 of Ellis et al.~\cite{2006APh....25..402E}, after excluding all short GRBs. These bursts were observed by the BATSE, HETE, and Swift satellites. The time delays were extracted by analyzing the spectral lags between the energy bands $115 \sim 320$ keV and $25 \sim 55$ keV using publicly available light curve data. The third subset comprises 21 time delays listed in Table~3 of Xiao et al.~\cite{Xiao:2022ovb}, which were obtained from GBM and Swift observations. These data correspond to spectral lags between the low-energy band of $15\sim 70$ keV and the high-energy band of $120 \sim 250$ keV, within a redshift range of $0.117 < z < 2.938$. 

\begin{figure}[h]
    \includegraphics[scale = 0.6]{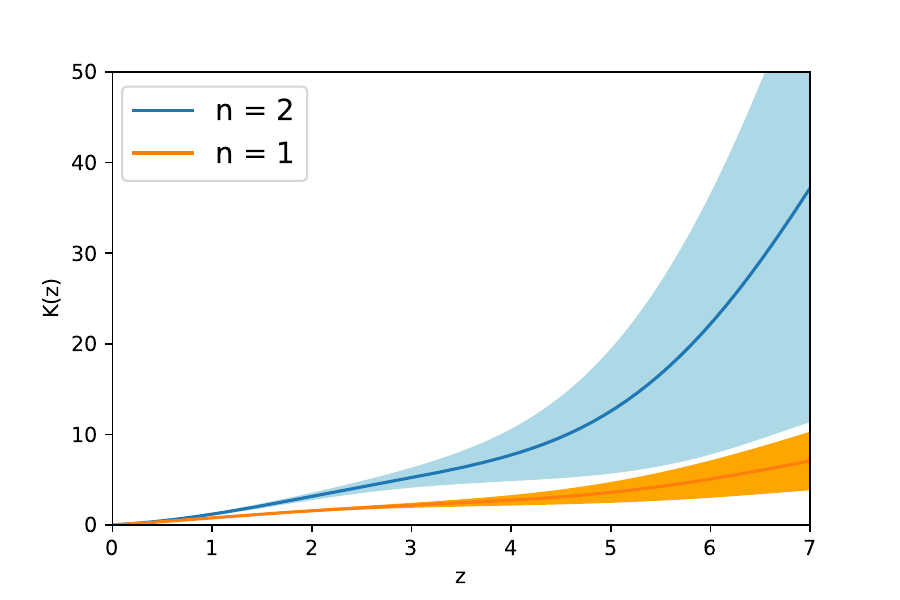}
    \caption{\label{fig:epsart} Reconstructed $K(z)$ as a function of redshift for $n=1$ (orange) and $n=2$ (blue). The shaded regions represent the $1\sigma$ uncertainties. }
\end{figure}

The 174 luminosity distance data used are calibrated using a cosmological model-independent method~\cite{Tang:2021elg}. These data are derived from a complete sample of gamma-ray and X-ray observations collected by the Swift satellite, covering a redshift range of $0.117<z<9.4$.
The luminosity distances are derived using the GRB correlation known as the Combo-relation~\cite{Izzo:2015vya}.
The Combo relation is given by
\begin{eqnarray}
\log(\frac{L_{0}}{\mathrm{erg/s}})=\log(\frac{A}{\mathrm{erg/s}})+\gamma \log(\frac{E_{p,i}}{\mathrm{keV}})\nonumber
\\-\log(\frac{\tau_{c}/s}{\mid1+\alpha_{c}\mid})
\label{eq:8},
\end{eqnarray}
where $\log$ denotes the base-10 logarithm; $L_{0}$ represents the isotropic equivalent luminosity during the plateau phase; and $\gamma$ and $A$ correspond to the slope and intercept parameters, respectively. $E_{p,i}$ is the peak energy of the $\nu F_{\nu}$ spectrum in the rest frame, while $\alpha_{c}$ and $\tau_{c}$ denote the decay index during the late power-law phase and the characteristic timescale at the end of the plateau, respectively. 

Given the luminosity distance of a GRB, the luminosity $L_{0}$ during the plateau phase can be derived from the observed energy flux $F_{0}$ in the rest-frame $0.3\sim10$ keV band as
\begin{eqnarray}
	L_{0}=4\pi D^{2}_{L}F_{0}
	\label{eq:9},
\end{eqnarray}
where the luminosity distance $D_{L}$ is related to the distance modulus $\mu$ through
\begin{eqnarray}
	\mu=5\log\frac{D_{L}}{\mathrm{Mpc}}+25
	\label{eq:10}.
\end{eqnarray}

The GRB distance modulus is given by~\cite{Izzo:2015vya,Muccino:2020gqt,Tang:2021elg}
\begin{align}
\mu_{\mathrm{GRB}}=-97.45+\frac{5}{2}[ \log(A)+\gamma \log(E_{p,i})\nonumber
 \\-\log(\frac{\tau_{c}}{|1+\alpha_{c}|})-\log(F_{0})-\log4\pi]
	\label{eq:11},
\end{align}
where $A$ and $\gamma$ are free parameters.

A neural network was trained using the Pantheon compilation of Type Ia supernovae and applied to calibrate 174 GRBs, producing the best-fit parameters $\gamma=0.856$, $\log A=49.661$, and $\delta_{\mathrm{GRB}}=0.228$~\cite{Muccino:2020gqt}. In our work, we bypass this calibration step and directly adopt these results.

\subsection{\label{32}Methods}

To implement a cosmological model-independent constraint on LIV, we replace the term $h(z)$ in $K(z)$ with an expression that does not rely on any specific cosmological model. The function $K(z)$ is then reconstructed using the luminosity distance data of GRBs. These luminosity distances are calibrated in a model-independent method by Tang et al.~\cite{Tang:2021elg}.
The expression for the luminosity distance in a flat universe can be written as
\begin{eqnarray}
	D_{L}(z) = \frac{c(1+z)}{H_{0}}\int_{0}^{z} \frac{dz^{\prime}}{h(z^{\prime})}
	\label{eq:12},
\end{eqnarray}
where c is the speed of light and $H_{0}$ is the Hubble constant, here we take $H_{0}=70$ km/s/Mpc.

After a simple transformation of the equation, the derivative of $K(z)$ can be expressed as
\begin{eqnarray}
	\frac{dK(z)}{dz} = \frac{(1+z)^{n}}{h(z)} = 
    \frac{d(\frac{H_{0}}{c(1+z)}D_{L})}{dz}(1+z)^{n}
    \label{eq:13},
\end{eqnarray}
the value of $dK(z)/dz$ has been calculated from $D_{L}$ given by the ‘Combo relation’. The derivative $dK(z)/dz$ is computed numerically from the discrete data, yielding a set of points distributed in redshift $z$. It is then numerically integrated to reconstruct the function $K(z)$ at discrete redshifts.

It should be noted that when applying both the time delay data and the reconstructed $K(z)$, the redshift $z$ of the two datasets must be matched.
Using the GP method~\cite{Seikel:2012uu}, the 174 scattered luminosity distance $D_{L}$ (obtained by 174 GRBs) can be extended to form a nearly continuous function of redshift $z$.
The GP reconstruction is entirely data-driven and independent of cosmological models.

The reconstructed $K(z)$ is shown in Fig.~\ref{fig:epsart}.
The $K(z)$ reconstruction is driven entirely by the luminosity distance data from the 174 GRBs and is independent of the cosmological model.

\begin{figure}[h]
	\includegraphics[scale = 0.35]{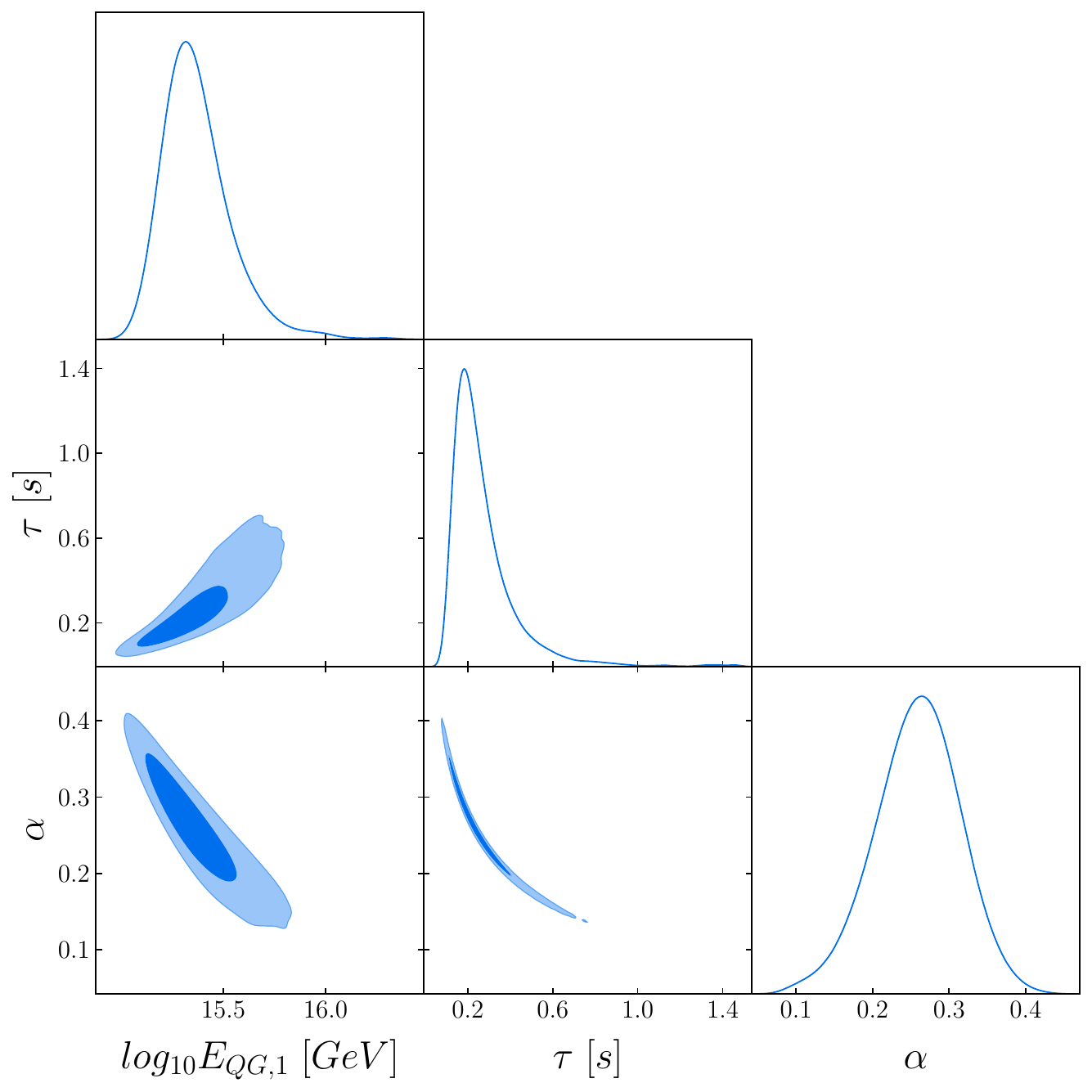}
	\caption{\label{fig:cc} The individual probability distribution of each parameter and the confidence contours in two dimensions for the parameters $log_{10}E_{QG}$, $\tau$, and $\alpha$ in 1$\sigma$ (considering linear LIV scenario with n = 1), the light is the 68\% contour, and the dark is the 95\% contour.}
\end{figure}

\section{\label{sec:results}Results and Discussion} 

In this paper, the Markov chain Monte Carlo (MCMC) method is employed to constrain the free parameters ($\log_{10}E_{\mathrm{QG},n},\tau,\alpha$). 
Its parameters range from [5, 30] GeV, [-10, 10] s, and [-10,10], with uniform prior probability distributions set within these ranges.
The constraints are derived by replacing the cosmological model with the reconstructed $K(z)$, and minimizing the chi-squared function, which is defined as
\begin{eqnarray}
\chi^{2}= \sum^{n}_{i = 1}  \left [  \frac{\Delta t_{\mathrm{obs}}-\Delta t_{\mathrm{th}}(E_{\mathrm{QG},n},\tau, \alpha)}{\sigma_{\mathrm{tot}}} \right ]^{2}
\label{eq:14},
\end{eqnarray}
where $\Delta t_{\mathrm{obs}}$ and $\Delta t_{\mathrm{th}}$ denote the observed and theoretical time delays for the GRBs, and $\sigma_{\mathrm{tot}}$ represents the corresponding total uncertainty. The latter is given by
\begin{eqnarray}
	\sigma_{\mathrm{tot}}^{2} = \sigma_{t}^{2}+(\frac{\partial f}{\partial E})^{2}\sigma_{E}^{2}+(\frac{\partial f}{\partial E_{0}})^{2}\sigma_{E_{0}}^{2}+\sigma_{D_{L}}^{2}
	\label{eq:15},
\end{eqnarray}
where $f$ represents the specific model; $\sigma_{t}$ indicates the uncertainty in spectral lag; $\sigma_{E_{0}}$ and $\sigma_{E}$ represent the widths of the lower and upper energy bounds, respectively; and $\sigma_{D_{L}}$ represents the uncertainty from the luminosity distance data.

To ensure convergence of results, we set the chain length to 300,000. This article investigates time delays in linear and quadratic relationships, corresponding to $n = 1$ and $n = 2$, respectively.
The fitting results with a confidence interval of 1$\sigma$ (68\%) are presented in Table~\ref{tab:table1}.

\begin{table}
\caption{\label{tab:table1}
	The 1$\sigma$ Bounds on $\log_{10}E_{\mathrm{QG},n}$, $\tau$, and $\alpha$ for Various LIV Models ($n = 1$, $n = 2$)
}
\begin{tabular}{llrrrrrrrrrrrlrrrrrrrrrrrrrlr}
\hline
\textrm{Model parameter}&
		\textrm{n = 1}&
		\textrm{n = 2} \\

\hline
    $\log_{10}E_{\mathrm{QG},n} \ (\mathrm{GeV})$ & $15.37_{-0.19}^{+0.10}$ & $9.99_{-0.06}^{+0.05}$ \\
		$\tau \ (\mathrm{s})$ & $0.26_{-0.15}^{+0.04}$ & $0.37_{-0.16}^{+0.07}$\\
		$\alpha$ & $0.26_{-0.05}^{+0.06}$ & $0.21_{-0.04}^{+0.04}$ \\
\hline
\end{tabular}
\end{table}

\subsection{LIV Energy Scale Constraints}

\begin{figure}[ht]
	\includegraphics[scale = 0.35]{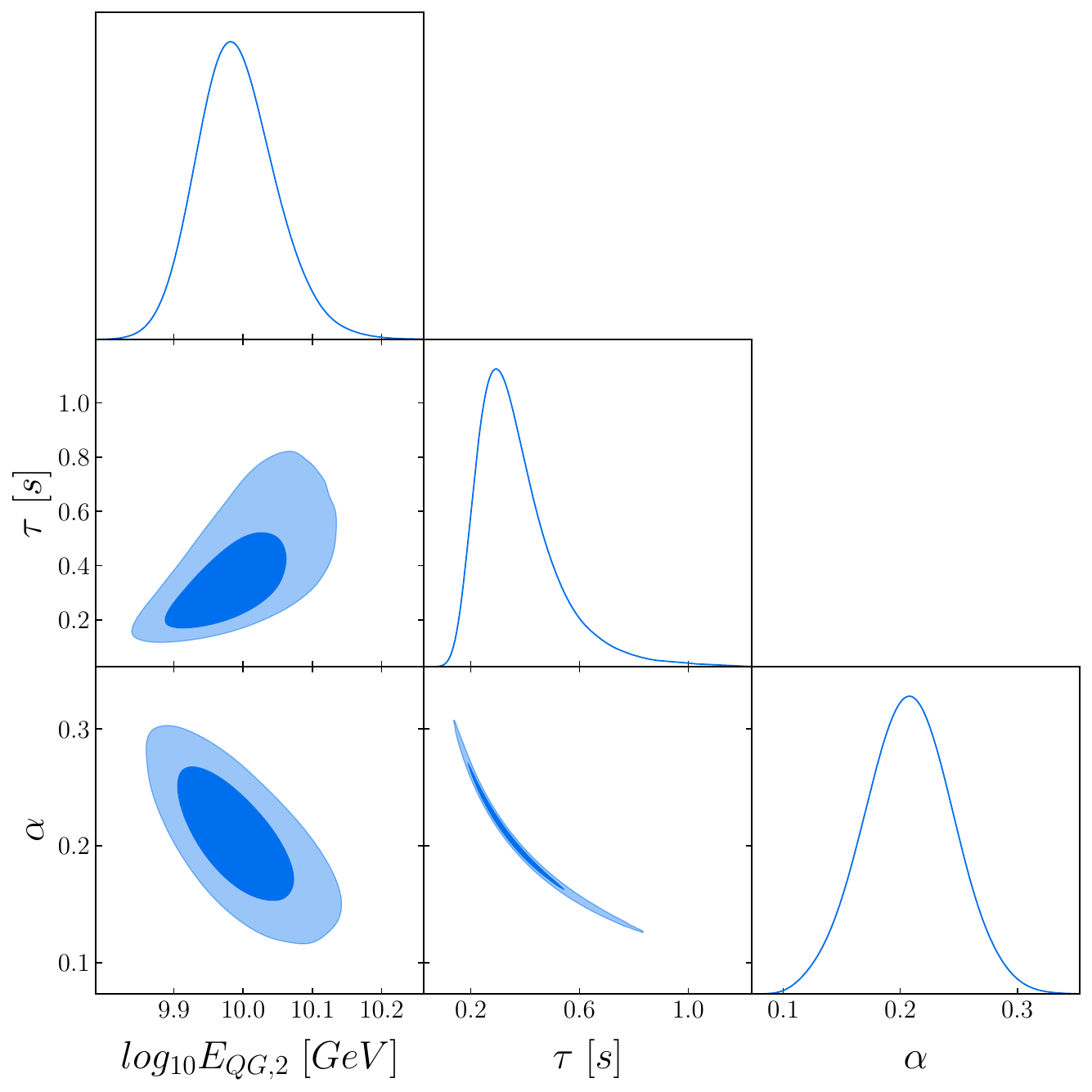}
	\caption{\label{fig:3} The individual probability distribution of each parameter and the confidence contours in two dimensions for the parameters $\log_{10}E_{QG}$, $\tau$, and $\alpha$ in 1$\sigma$ (considering quadratic LIV scenario with n = 2), the light is the 68\% contour, and the dark is the 95\% contour.}
\end{figure}

We initially fitted the lag energy in 1$\sigma$ under linear conditions ($n=1$).
Fig.~\ref{fig:cc} is a two-dimensional density plot illustrating the fitting results when $n = 1$.
And the best-fit parameter values are $\log_{10}E_{\mathrm{QG},1} = 15.37_{-0.19}^{+0.10} \ \mathrm{GeV}$, $\tau = 0.26_{-0.15}^{+0.04} \ \mathrm{s}$ and $\alpha=0.26_{-0.05}^{+0.06}$.
Then we consider the quadratic case in 1$\sigma$ which $n = 2$.
The parametric results of the constraints are shown in Fig.~\ref{fig:3}.
The best-fit parameter values are $\log_{10}E_{\mathrm{QG},2} =9.99_{-0.06}^{+0.05} \ \mathrm{GeV}$, $\tau=0.37_{-0.16}^{+0.07} \ \mathrm{s}$ and $\alpha = 0.21_{-0.04}^{+0.04}$.

With the best-fit values of $\log_{10}E_{\mathrm{QG},1}$ and $\log_{10}E_{\mathrm{QG},2}$ as well as their $1\sigma$ error bars, the $1 \sigma$ confidence-level lower limit on LIV is $E_{\mathrm{QG},1}\geq 1.5\times 10^{15}$ GeV and $E_{QG,2}\geq 8.5\times 10^{9}$ GeV.
The $E_{\mathrm{QG},1}$ is reduced by four orders of magnitude compared to the Planck energy ($E_{\mathrm{Pl}} \sim 1.22 \times 10^{19} \mathrm{GeV} $), and the energy $E_{\mathrm{QG},2}$ is significantly smaller than the Planck energy by a factor of ten orders of magnitude.
In the linear case ($n=1$), our results are slightly lower than those obtained in studies based on 35 time delays~\cite{Wei:2016exb} and 37 time delays~\cite{Ellis:2005sjy}.
However, for the quadratic case ($n=2$), our result is nearly three orders of magnitude higher than their results.
Compared with the GP method employed by Pan et al.~\cite{Pan:2020zbl}, the obtained lower bound of $E_{\mathrm{QG},1}$ in the linear scenario is elevated by an order of magnitude, and the result in the quadratic scenario is also raised by approximately an order of magnitude.
Specifically, compared to the analysis based on a pure short GRB sample by Xiao et al.~\cite{Xiao:2022ovb}, which reported $E_{\mathrm{QG},1} \geq 2.5 \times 10^{15}~\mathrm{GeV}$ and $E_{\mathrm{QG},2} \geq 1.2 \times 10^{6}~\mathrm{GeV}$ under the assumption of constant intrinsic time delays, our linear-order constraint $(n = 1)$ is very close to their result.
In terms of constraint accuracy, our quadratic result is significantly higher than Wei et al.~\cite{Wei:2016exb}, but the linear result is lower than their results.
Although our results are somewhat lower in precision than the previous ones, this approach may potentially circumvent unknown discrepancies among various cosmological probes.

\subsection{Intrinsic Time Delay Constraints}

The power-law model of intrinsic time delay provides a physically motivated description of GRB emission and can be regarded as an empirical framework for characterizing spectral lags.
For the case of $n = 1$, the best-fit parameters are $\tau = 0.26^{+0.04}_{-0.15} \ \mathrm{s}$ and $\alpha = 0.26^{+0.06}_{-0.05}$.  
When $n = 2$, we obtain $\tau = 0.37^{+0.07}_{-0.16} \ \mathrm{s}$ and $\alpha = 0.21^{+0.04}_{-0.04}$. 
In both cases, within the $1\sigma$ confidence interval, the best-fit values satisfy $\tau > 0$ and $\alpha > 0$, indicating a positive intrinsic delay---i.e., higher-energy photons arrive earlier than lower-energy ones when only intrinsic source effects are considered. This implies that, in the source rest frame, higher-energy emission likely precedes lower-energy emission.

The positive intrinsic delay has been predicted or discussed in several GRB emission scenarios.
For example, in synchrotron internal-shock models~\cite{2014A&A...568A..45B}, the spectral peak energy $E_{\mathrm{peak}}$ decreases with time due to radiative cooling of relativistic electrons.
As a result, emission at higher energies peaks earlier than that at lower energies, producing the commonly observed hard-to-soft spectral evolution within pulses.
Positive intrinsic delay may also arise from kinematic and geometric effects, such as relativistic curvature~\cite{2009ApJ...703.1696Z} or jet-structure models~\cite{Qin_2004}.
In these scenarios, photons emitted at different angles relative to the line of sight experience different light-travel times and Doppler boosting.
As a result, even emission that is symmetric in the comoving frame can appear asymmetric and energy dependent in the observer frame, with lower-energy photons arriving later.
Magnetic dissipation models provide another possible explanation, in which gradual energy release and particle acceleration can lead to systematic delays between photons of different energies~\cite{2011ApJ...726...90Z}.

Beyond interpretations based on a single prompt emission episode, an alternative scenario attributes the positive intrinsic delay to a preburst emission phase at high energies~\cite{2021PhLB..82036518Z}. Systematic analyses of large GRB samples have shown that, in a non-negligible fraction of events, a preburst phase is present in which high-energy photons are emitted earlier at the source than the onset of the main prompt emission~\cite{2016APh....82...72X,2016arXiv160708043X,2021PhLB..82036546Z,2021JHEAp..32...78C}. When high-energy photons produced during the preburst phase are compared with lower-energy photons emitted later during the prompt phase, a positive delay naturally emerges. Observations of GRB~221009A also provide support for this scenario, as high-energy photons were directly detected before the main prompt phase, indicating the presence of an early high-energy emission component preceding the conventional prompt emission~\cite{2024RNAAS...8..263L}.

From a statistical perspective, long GRBs population studies show that positive spectral lags are prevalent~\cite{2017ApJ...844..126S}. The dominance of positive lags in long GRB samples is therefore consistent with the positive intrinsic delay parameters inferred in our analysis, lending further support to the adopted intrinsic delay model.

\subsection{Intrinsic Delay Model Comparison}

Long GRBs exhibit a broad range of intrinsic emission timescales, from a few seconds to several hundred seconds, and the physical origin of these delays remains uncertain~\cite{2009ARA&A..47..567G}. Therefore, modeling the intrinsic emission delay is essential. Early studies often adopted a common intrinsic delay term, implicitly assuming that all GRBs share the same intrinsic delay.
While this assumption provides a simple baseline description, it is less well suited for long GRBs, which exhibit a wide diversity of intrinsic emission timescales.

A more flexible and widely used approach models the intrinsic delay as an energy-dependent function. The power-law model of Eq.~\ref{eq:6} is motivated by empirical studies of the spectral evolution observed in large GRBs (mostly long GRBs)~\cite{2017ApJ...844..126S}. This form is capable of describing multi-band (keV--MeV) spectral lags measured from GRB light curves, where the energy channels are well sampled~\cite{Wei:2016exb}. The parameters $\tau$ and $\alpha$ respectively describe the amplitude and energy dependence of the intrinsic delay and allow for both high- and low-energy leading behavior.

An alternative parameterization assumes a linear dependence of the intrinsic delay on the photon energy in the source frame~\cite{2024PhLB..85638951S}. This model consists of two components: a common constant offset $\Delta t_c$ and an energy-dependent term proportional to the source-frame energy $E_s$ of the high-energy photon. The motivation is based on empirical indications of an energy-dependent emission structure among photons with energies above the GeV scale. The intrinsic delay is written as
\begin{equation}
\Delta t_{\rm int} = \tau E_s + \Delta t_c,
\end{equation}
where $\tau$ quantifies the correlation between emission time and photon energy, and $\Delta t_c$ represents a common intrinsic offset shared among events.

Both parameterizations are phenomenological and assume an energy-dependent intrinsic emission time to account for unknown source-intrinsic effects. They allow for both positive and negative intrinsic delays. In both approaches, broad and uninformative priors are adopted for the intrinsic-delay parameters to avoid biasing the LIV constraints. The primary difference lies in the treatment of the energy dependence. The linear model places greater weight on high-energy photons and suppresses the contribution from lower-energy emission. It is therefore well suited for very-high-energy GRBs and preburst scenarios. By contrast, the power-law model incorporates photons over a wide energy range and allows different energy bands to contribute comparably to the overall delay. As a result, it is more sensitive to geometric and spectral-evolution effects, and is better suited for describing the commonly observed temporal evolution of GRB light curves, where emissions in different energy bands occur at systematically different times.

In our analysis, the relevant photon energies lie in the keV--MeV range, with the majority of measurements concentrated in the keV band. The time-delay measurements involve multiple energy bands within individual GRBs. The energy separation between the low- and high-energy photons is moderate, and thus the contribution from lower-energy photons cannot be ignored. In this energy regime, GRB spectral evolution often exhibits non-linear behavior that cannot be adequately described by a linear dependence on photon energy~\cite{2018ApJ...869..100U}. For these reasons, a power-law intrinsic delay model motivated by multi-band keV--MeV GRB observations is more appropriate for our analysis. 

Nevertheless, it should be emphasized that intrinsic delay models currently used in GRB studies remain phenomenological~\cite{1998Natur.395Q.525A,2021ApJ...906....8D}. Different parameterizations, together with their associated prior assumptions, can lead to degeneracies between intrinsic emission delays and propagation effects, thereby rendering the inferred LIV constraints model dependent~\cite{2024PhLB..85638951S,2025PhRvD.111j3015S,2023PhRvD.108l3023V}. A more physically motivated description of intrinsic delays would help reduce these degeneracies and is expected to lead to more reliable and potentially tighter constraints on LIV.

\section{\label{sec:conclusion}conclusion}

In this study, we apply a new cosmological model-independent restriction to LIV using time delay data and luminosity distance data from GRBs.
Initially, we collected a dataset consisting of 88 time delay obtained from long GRBs, reaching the redshift to $z\sim 6.29$. Subsequently, we employ the GP methodology to reconstruct the luminosity distance datasets from 174 GRBs into $K(z)$. Additionally, the power-law model of intrinsic time delays is incorporated in our LIV model, with two parameters $\tau$ and $\alpha$.

Under the constraints of multiband GRBs, we get the LIV limit results at higher redshifts ($0.117<z<6.29$), the linear and quadratic LIV effects are $E_{\mathrm{QG},1}\geq 1.5\times 10^{15}$ GeV and $E_{\mathrm{QG},2}\geq 8.5\times 10^{9}$ GeV. 
The magnitude exhibits a significant reduction in comparison to the energy scale of Planck.
With the support of the 88 time delay dataset from long GRBs, we obtain the relatively accurate intrinsic time delay constraint:
$\tau = 0.26_{-0.19}^{+0.10}$ s and $\alpha=0.26_{-0.05}^{+0.06}$ ($n=1$), $\tau=0.37_{-0.16}^{+0.07}$ s and $\alpha = 0.21_{-0.04}^{+0.04}$ ($n=2$).
Our results show a positive intrinsic time delay, indicating that higher-energy photons are emitted prior to lower-energy ones.
All of the results are consistent with the previous studies.
We combine the luminosity distance with the time delay information used in this paper, which comes entirely from GRBs.
This might alleviate some systematics coming from different physical properties of different populations of objects, providing a more compelling set of constraints.

In summary, we present a novel approach to constrain LIV independent of cosmological models.
To achieve tighter LIV constraints, the support from a larger number of GRBs time delay data with enhanced accuracy and higher energy band is anticipated in our study. Combined with more observations, we may be able to further investigate the constraints of LIV and explore the underlying physics behind LIV.

\begin{acknowledgments}
This work was supported by the National Natural Science Foundation of China Grant No. 12105032. Shuo Cao is partially supported by Beijing Natural Science Foundation No. 1242021; the National Natural Science Foundation of China (Nos. 12203009, 12433001).
\end{acknowledgments}

\nocite{*}

\bibliography{apssamp}

\end{document}